\documentclass[3p,times,twocolumn]{elsarticle}

\usepackage{amssymb}
\usepackage{amsthm}
\usepackage{amsmath}
\usepackage[ruled,vlined]{algorithm2e}
\usepackage{float}
\usepackage{lineno} 
\usepackage{xcolor}
\begin{document}
\begin{frontmatter}

\title{Use of Bayesian Optimization to Understand the Structure of Nuclei}

\author[a]{J.~Hooker}
\author[a]{J.~Kovoor}
\author[a]{K.L.~Jones\corref{cor1}}
\cortext[cor1]{Department of Physics and Astronomy, 401 Nielsen Physics Building, 1408 Circle Drive, University of Tennessee, Knoxville, TN 37996, USA }
\ead{kgrzywac@utk.edu}
\author[b,c]{R.~Kanungo}
\author[c]{M.~Alcorta}
\author[d]{J.~Allen}
\author[e]{C.~Andreoiu}
\author[f]{L.~Atar}
\author[d]{D.W.~Bardayan}
\author[c]{S.S.~Bhattacharjee  \fnref{label1}}
\fntext[label1]{Current address Institute of Experimental and Applied Physics, Czech Technical University in Prague 110 00 Prague 1, Czech Republic}
\author[d]{D.~Blankstein}
\author[f]{C.~Burbadge}
\author[a]{S.~Burcher}
\author[g]{W.N.~Catford}
\author[h]{S.~Cha}
\author[h]{K.~Chae}
\author[c]{D.~Connolly}
\author[c]{B.~Davids}
\author[c]{N.~Esker}
\author[e]{F.H.~Garcia}
\author[c]{S.~Gillespie}
\author[a]{R.~Ghimire}
\author[d]{A.~Gula}
\author[c]{G.~Hackman}
\author[g]{S.~Hallam}
\author[b]{M.~Hellmich}
\author[c]{J.~Henderson}
\author[b,c]{M.~Holl}
\author[b]{P.~Jassal}
\author[i]{S.~King}
\author[b]{T.~Knight}
\author[c]{R.~Kruecken}
\author[j]{A.~Lepailleur}
\author[i]{J.~Liang}
\author[g]{L.~Morrison}
\author[d]{P.D.~O'Malley}
\author[k,a]{S.D.~Pain}
\author[a]{X.~Pereira-Lopez}
\author[i]{A.~Psaltis \fnref{label2}}
\fntext[label2]{Current address Institut f\"ur Kernphysik, Technische Universit\"at Darmstadt, Darmstadt 64289, Germany}
\author[f]{A. Radich}
\author[l]{A.C.~Shotter}
\author[a]{M.~Vostinar}
\author[c]{M.~Williams}
\author[b]{O.~Workman}

\address[a]{Department of Physics and Astronomy, University of Tennessee, Knoxville, Tennessee 37996, USA}
\address[b]{Astronomy and Physics Department, Saint Mary's University, Halifax, Nova Scotia B3H 3C3, Canada}
\address[c]{TRIUMF, Vancouver, British Columbia V6T 4A3, Canada}
\address[d]{Physics Department, University of Notre Dame, Notre Dame, Indiana 46556, USA}
\address[e]{Department of Chemistry, Simon Fraser University, Burnaby, British Columbia, V5A 1S6, Canada}
\address[f]{Department of Physics, University of Guelph, Guelph, Ontario N1G 2W1, Canada}
\address[g]{Department of Physics, University of Surrey, Guildford GU2 7XH, UK}
\address[h]{Department of Physics, Sungkyunkwan University, 2066 Seobu-ro, Jangan-gu, Suwon, Korea}
\address[i]{Department of Physics \& Astronomy, McMaster University, 1280 Main St. W, Hamilton, Ontario L8S 4M1, Canada}
\address[j]{Department of Physics and Astronomy, Rutgers University, New Brunswick, New Jersey 08903, USA}
\address[k]{Physics Division, Oak Ridge National Laboratory, Oak Ridge, Tennessee 37831, USA}
\address[l]{School of Physics and Astronomy, University of Edinburgh, JCMB, Mayfield Road, Edinburgh EH9 3JZ, United Kingdom}

\date{\today}

\begin{abstract}
Monte Carlo simulations are widely used in nuclear physics to model experimental systems.   In cases where there are significant unknown quantities, such as energies of states, an iterative process of simulating and fitting is often required to describe experimental data.  We describe a Bayesian approach to fitting experimental data, designed for data from a $^{12}$Be(d,p) reaction measurement, using simulations made with GEANT4. Q-values from the $^{12}$C(d,p) reaction to well-known states in $^{13}$C are compared with simulations using BayesOpt.  The energies of the states were not included in the simulation to reproduce the situation for $^{13}$Be where the states are poorly known.  Both cases had low statistics and significant resolution broadening owing to large proton energy losses in the solid deuterium target.  Excitation energies of the lowest three excited states in $^{13}$C were extracted to better than 90~keV, paving a way for extracting information on $^{13}$Be.   
\end{abstract}

\end{frontmatter}


\section{Introduction}
One of the main goals in nuclear physics is to expand the limits of observation of nuclear structure through reactions involving exotic nuclei.
Direct reactions, such as the (d,p) one-neutron transfer reaction, have been used extensively to study the structure of nuclei. 
With the limited beam time available at radioactive ion beam facilities, and the associated cost of running these experiments, it is incumbent on experimentalists to extract the maximum information from the data obtained.  There is also a need to account for sources of background and assess uncertainties in experimental data.
Simulations can be used to understand the experimental resolution, which is commonly a combination of interactions of the beam and final-state particles with both the target and other materials such as detectors. 
They can also account for the effects of incomplete acceptances, which can complicate data analyses.

The GEANT4 toolkit is commonly used to simulate the passage of particles through matter by taking into account both electromagnetic and nuclear processes using Monte Carlo (MC) methods \cite{Agostinelli:2002}.
GEANT4 provides a general framework for MC simulations of particles with support for detector construction, particle transportation, source generation, and particle detection.
These simulations allow the user to understand the detector response for their specific conditions and reactions for a given experiment.
This paper reports on the use of GEANT4 in combination with Bayesian Optimization for understanding reaction data and thereby extracting nuclear structure information.\\
A study using the $^{12}$C(d,p)$^{13}$C transfer reaction was performed immediately before the experiment to study the structure of the unbound nucleus $^{13}$Be via the $^{12}$Be(d, p) reaction and was used to check the detectors and provide calibration points.
As $^{12}$Be has a half life of 21~ms \cite{Nudat:2021}, the experiment necessarily was run in inverse kinematics, with a beam of $^{12}$Be and a solid deuterium target, as discussed in detail below in Section \ref{sec:exp}.  The $^{12}$C(d,p) reaction was run in inverse kinematics to mimic the reaction with the radioactive ion beam, and both had much lower statistics than usually obtained with stable ion beams.  The $^{12}$C beam had to be reduced in intensity to avoid damaging the forward angle silicon detectors.  \\
The nucleus $^{13}$C has a very well-known structure and provides an excellent benchmark for testing these methods.
The first four states in $^{13}$C have been observed several times through $^{12}$C(d, p)$^{13}$C transfer reactions with excitation energies of: $0$, $3.089$, $3.685$, and $3.854$ MeV \cite{Nudat:2021}.
Here we describe the benchmarking of a GEANT4 simulation of the experimental setup with Bayesian Optimization (discussed in Section \ref{sec:BayesOpt}) using experimental data to extract the underlying nuclear structure.

\section{Experiment and Simulation Description}
\label{sec:exp}

The $^{12}$C(d, p) reaction was performed using the IRIS facility at the ISAC II experimental area of TRIUMF.
IRIS was designed for the study of nucleon transfer reactions and inelastic scattering of exotic nuclei in inverse kinematics \cite{Kanungo:2014}.
The central component of the IRIS facility is a thin, solid hydrogen, or deuterium, target, which avoids the significant amounts of carbon present in polyethylene foils (CH$_{2}$)$_{\text{n}}$ and (CD$_{2}$)$_{\text{n}}$, which are commonly used in direct reactions measurements.  The carbon in CD$_2$ targets can create fusion evaporation background and is a large source of energy loss and thereby uncertainty in energy.
To create the solid deuterium target, a thin Ag foil (4.64 $\mu$m) is cooled via a helium compressor to a temperature around $\sim$4K.
The target gas (hydrogen or deuterium) is then sprayed onto the foil where the gas solidifies.
The thickness of the target is controlled by adjusting the amount of gas sprayed on the foil.\\
\begin{figure}
    \centering
	\includegraphics[scale=0.32, angle=-90]{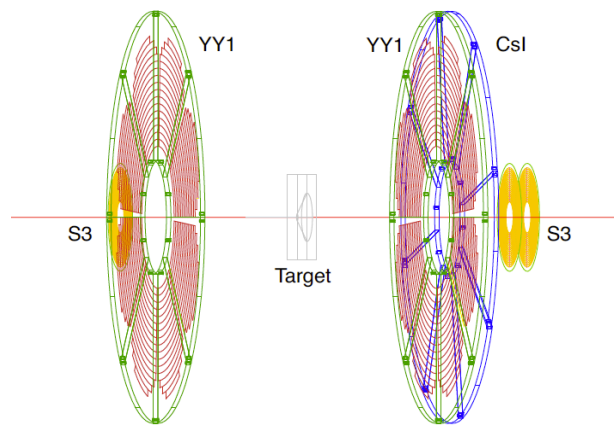}
    \caption{Experimental setup as represented in GEANT4. 
    Two YY1 detectors, one located upstream and one downstream from the target were used to measure the ejected protons from the (d, p) reaction. 
    Downstream from the second YY1 detector are two S3 detectors to measure the heavy recoil.
             \label{fig:ExperimentalSetup}}
\end{figure}
The IRIS facility incorporates a suite of detectors upstream and downstream from the cryogenic target.  The specific configuration used in this experiment included an ionization chamber (IC), upstream of the target, filled with 19.4~Torr of isobutane (C$_{4}$H$_{10}$) to measure the energy losses of beam particles, providing beam particle identification.
The entrance and exit windows of the IC were 30 $\mu$m and 50 $\mu$m thick silicon nitride (SiN$_{3}$), respectively.
Annular silicon detectors (MICRON Semiconductor YY1 type) faced the target from the upstream and downstream sides and measured the angles and energies of light reaction particles.
The YY1 detectors have 8 azimuthal detector sectors where each detector is segmented into 16 rings.
The upstream YY1 detector, with a thickness of 500 $\mu$m, was placed 80.8 mm from the target, while the downstream YY1 detector with the 100~$\mu$m thickness was placed 86 mm from the target.
The downstream YY1 detector was backed by a 12~mm-thick cesium-iodide (CsI) scintillator forming a telescope for reaction particles emitted in the forward direction.
The telescope provides dE-E particle identification for particles that do not stop in the silicon detector.
Located 600 mm and 690 mm downstream from the target were two smaller annular silicon detectors (MICRON S3 type) used for dE-E particle identification of the heavy recoil.
The transmission S3 detector (dE) had a thickness of 61 $\mu$m, the S3 stopping (E) detector, further downstream, had a thickness of 500 $\mu$m.
The IRIS facility is described in more detail in Ref \cite{Kanungo:2014}.

A $111.4 \pm 2.2$ MeV beam of $^{12}$C at a rate of 1.5 x 10$^3$~pps impinged on the solid deuterium target, which had an average thickness of 56 $\mu$m.

\begin{figure}
    \centering
    \includegraphics[scale=0.30]{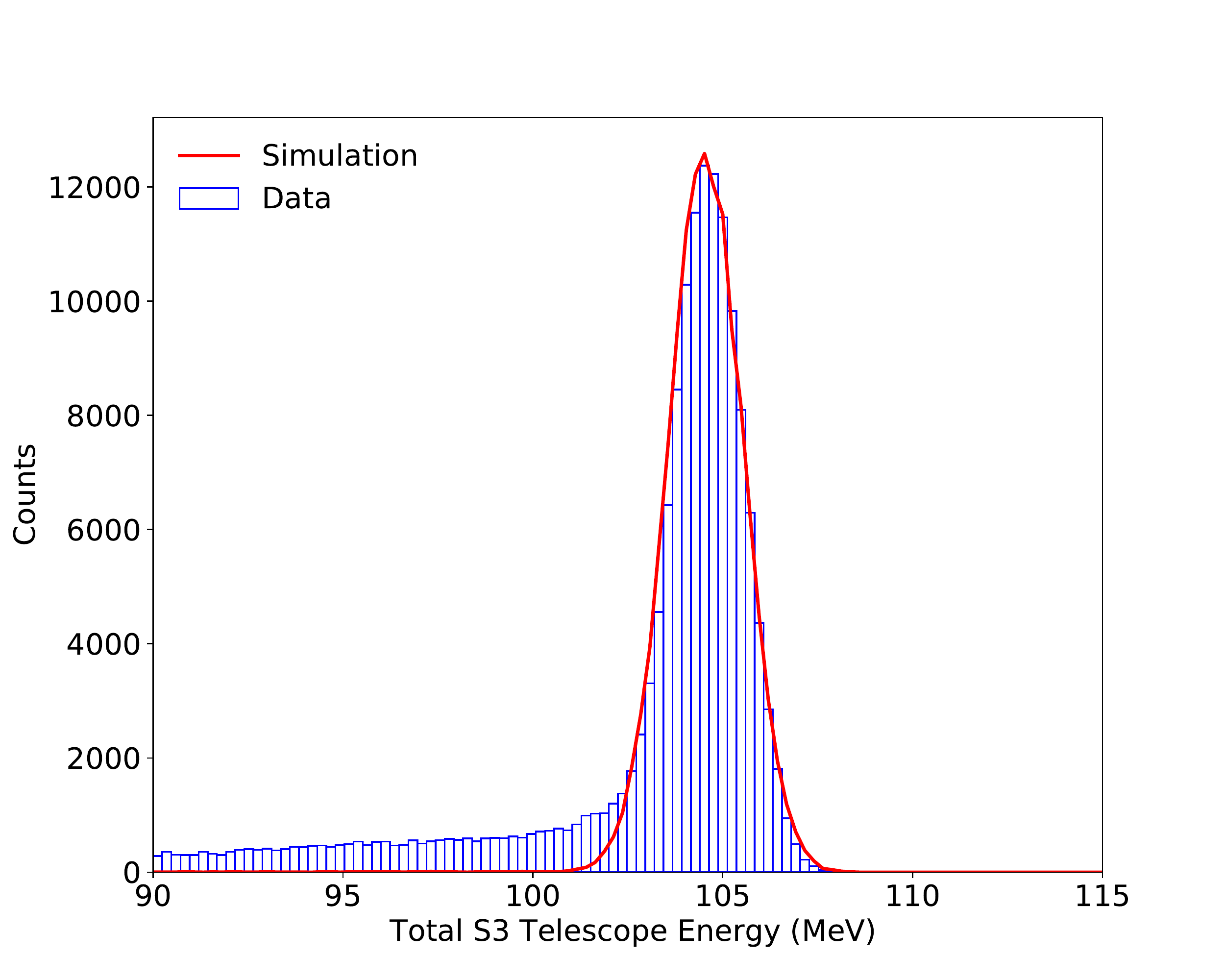}
    \caption{(Color online) Total energy measured in the S3 detectors with no solid deuterium target as simulated (red curve) compared with that measured in the experiment (blue bins).
             \label{fig:Beam_Data_Sim_Compare}}
\end{figure}

The experimental setup was reproduced in GEANT4 as shown in Figure \ref{fig:ExperimentalSetup}.
 The beam energy reproduction was tested in the simulation by comparison with data without the solid deuterium target.
A short run with the $^{12}$C beam and no target was performed, where only the silver foil was in the target location. The total energy of the $^{12}$C ions was measured in the S3 telescope.
The comparison of the data from the S3 detectors and the simulation is shown in Figure \ref{fig:Beam_Data_Sim_Compare}.
The simulation reproduces the peak energy seen in the S3 telescope. The low-energy tail below the 102-MeV peak was due to incomplete charge collection in the silicon detector as also observed in alpha source calibrations.

Only the upstream YY1 detector was considered for the measurement of the (d,p) reaction.
As the light reaction products from (d,d) and (d,t) reactions in inverse kinematics are constrained to the forward hemisphere in the laboratory frame, these reactions cannot be measured in the upstream YY1 detector.  The only direct reaction products that can be detected at backward angles in this experiment are protons from the (d,p) reaction.
Based on alpha calibrations, these detectors had an intrinsic FWHM resolution of 35 keV, which was implemented in the GEANT4 simulation.
The energy and angle of the proton, as measured in the YY1 detector, were used to reconstruct the Q value of the reaction.
Internal GEANT4 energy loss tables for the proton and the $^{12}$C beam in the solid deuterium target were extracted and used to calculate energy losses. The Q value was reconstructed from the detected energies, assuming the reaction occurred at the center of the target.
For consistency in the comparison of the data and simulation, the analysis of the experimental data also used the internal GEANT4 energy-loss tables for reconstructing the reaction Q-value.

\begin{figure}
    \centering
    \includegraphics[scale=0.30]{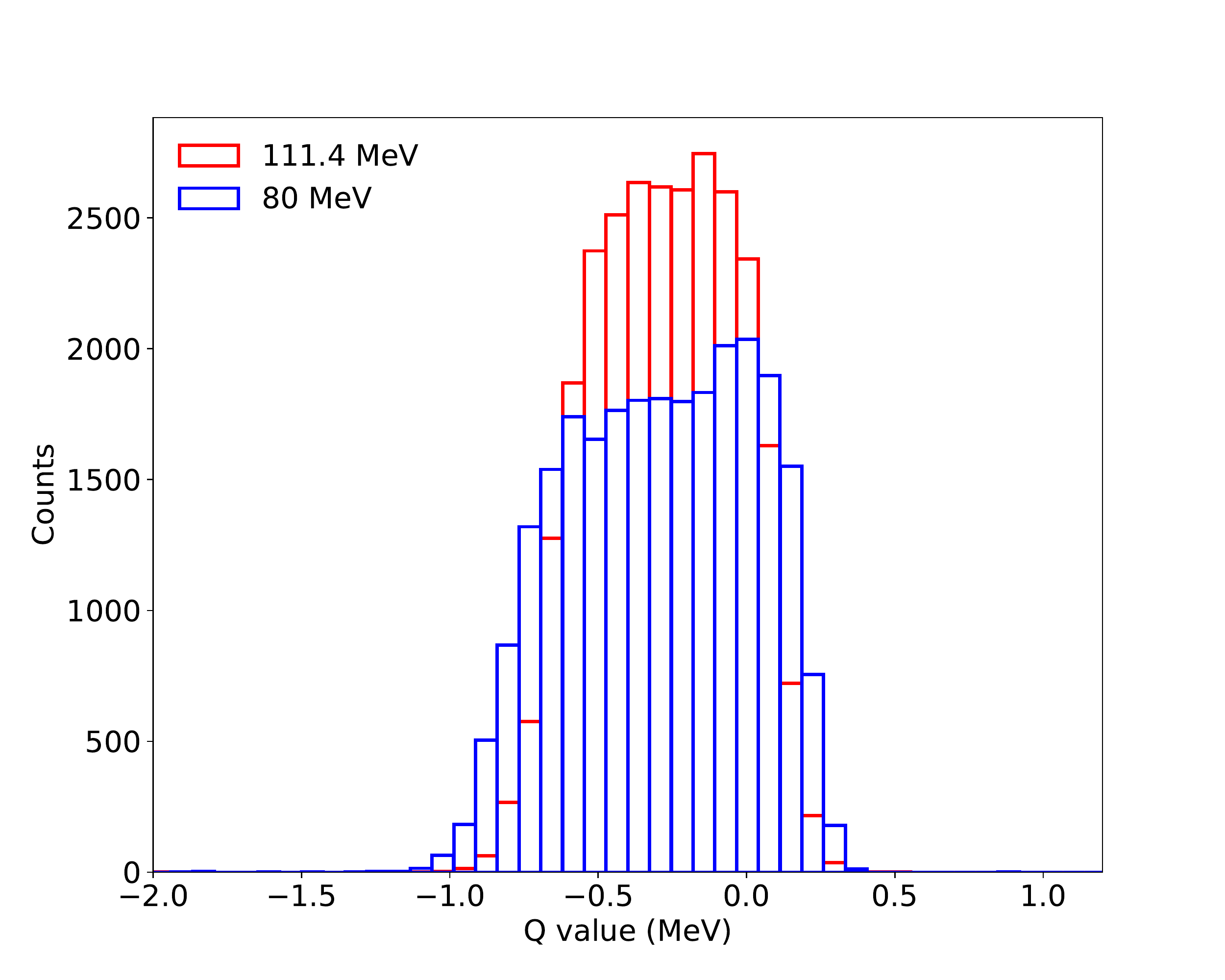}
    \caption{(Color online) Q value spectrum for the $^{12}$C(d, p) reaction simulated with GEANT4 for a hypothetical state at 3.0 MeV.
                            The red histogram is for a beam energy of 111.4~MeV corresponding to proton energies in the most backward angle of the YY1 detector of 1.79~MeV.
                            The blue histogram shows the same spectrum but for a beam energy of 80~MeV with proton energies in the most backward angle of the YY1 detector of 1.25~MeV.
                            The red histogram shows a Gaussian distribution while the blue histogram has a somewhat skewed detector response function.
             \label{fig:Resolution_Comparison}}
\end{figure}

The experimental resolution, which is dominated by the intrinsic detector resolutions, the energy straggling of the beam, and the energy straggling of the emergent light ion, is sometimes assumed to be Gaussian, and is included in fits as such.  This approach does not work in cases with low beam energy and energy-dependent resolutions, as demonstrated in the example discussed below and shown in Figure \ref{fig:Resolution_Comparison}. 
The $^{12}$C(d,p) reaction was simulated for a hypothetical state at an excitation energy of 3.0~MeV (Q value of -0.278~MeV).
The red histogram shows the Q-value spectrum from the simulation at the nominal beam energy of 111.4~MeV.
At this energy, the average proton energy for the most backward ring in the YY1 detector is 1.79~MeV.
The blue histogram shows the Q-value spectrum from a similar simulation with a beam energy of 80 MeV, where the average proton energy of the most backward ring in the YY1 detector is 1.25~MeV.
At lower proton energies, the resolution of the detected proton worsens owing to the larger energy loss, as shown in Figure \ref{fig:Resolution_Comparison} as the blue histogram is much wider than the red histogram.
The second effect occurs as the total energy and proton energy decrease, the detector resolution when extracting the Q value changes from a Gaussian distribution to a non-Gaussian distribution.
The Q value for these lower energies becomes skewed towards higher Q values, that is high proton energies at the center of the target.
This effect comes from protons generated at the furthest downstream position in the target having such a low energy that they either do not make it out of the target, or do not have enough energy when they hit the detector to make it above the noise threshold.
The proton energy at the center of the target is reconstructed by adding on the energy losses in the target and the dead layer of the detector. The Q value in the case of a (d,p) reaction is subsequently found using equation \ref{eqn:Q}.
\begin{equation}
    Q = \frac{M_{p} + M_{r}}{M_{r}}E_{p} - \frac{M_{r} - M_{b}}{M_{r}}E_{b} - \frac{2 \sqrt{M_{b}M_{p}E_{b}E_{p}}}{M_{r}} \cos{\theta}
    \label{eqn:Q}
\end{equation}
where $M_{b}$, $M_{p}$, and $M_{r}$ are the masses of the beam, emergent proton, and heavy recoil respectively, $E_{b}$ is the beam energy at the center of the target, $E_{p}$ is the proton energy at the center of the target, and $\theta$ is the angle of the light recoil particle.
The analysis of the simulated data is performed in the same way as for the experimental data to allow direct comparisons.

\begin{figure*}[ht]
\centering
\includegraphics[scale=0.3]{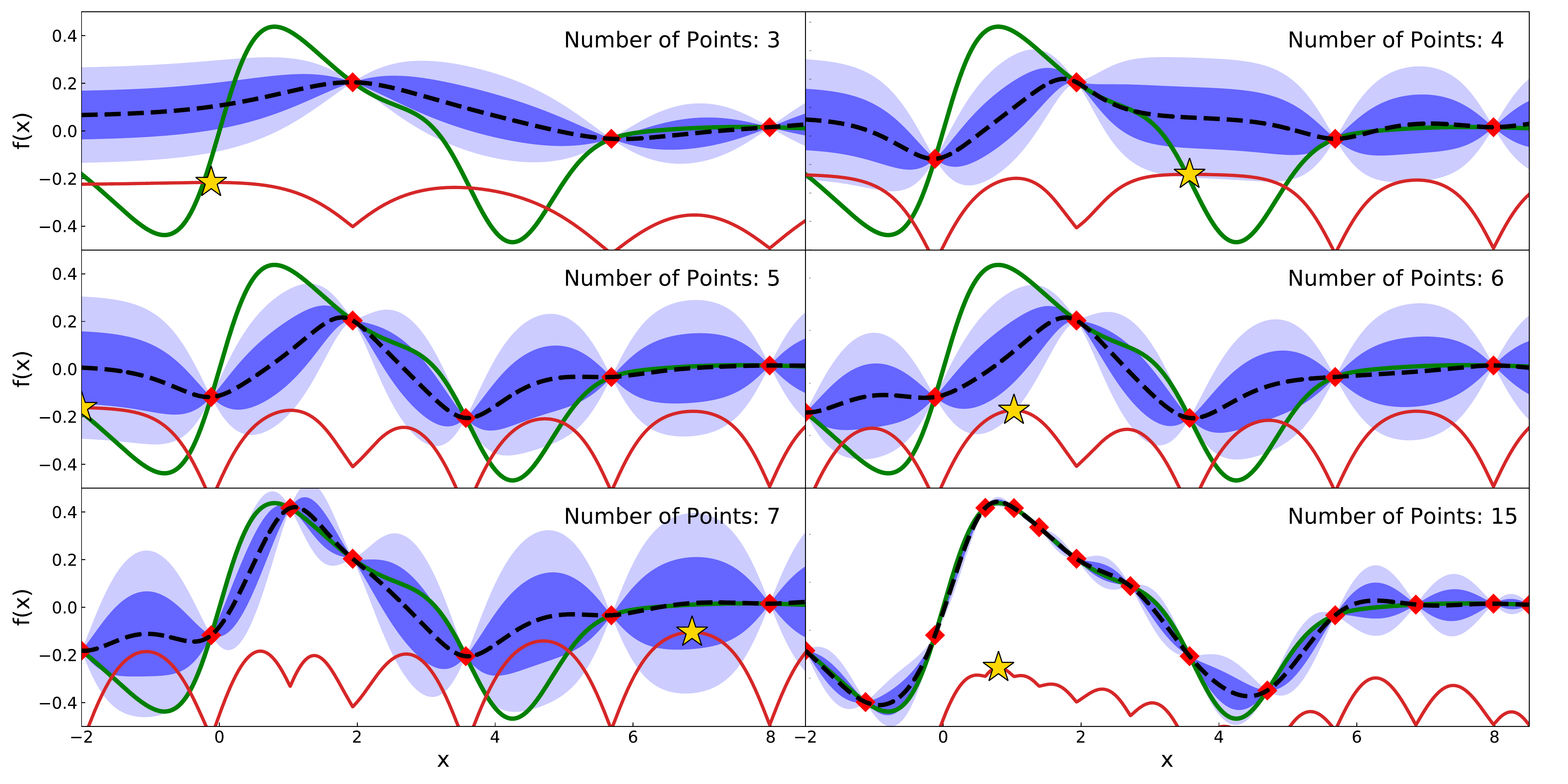}
\caption{A series of BayesOpt iterations to find the maximum of the test function $f(x) = \left( 1/(x^{2} + 1) + e^{-(x-4)^2/2} \right) \sin(x)$ (green solid line) over the range $x \in [-2, 8.5]$.
         In the first panel, three randomly selected points were used to evaluate the function shown as red filled markers.
         The acquisition function for the expected improvement is shown by the red solid line and the next point for evaluation by a yellow star.
         The mean and the 66\% and 95\% confidence intervals of the $\mathcal{GP}$ are shown by the black dashed line, and the dark and light blue shaded regions, respectively.
         The following panels show subsequent iterations in the Bayesian Optimization process, where the last panel shows the algorithm honing in on the maximum value of f(x) around $x=0.6$.
         \label{fig:BayesOptExample}}
\end{figure*}
\section{Bayesian Optimization} \label{sec:BayesOpt}

Bayesian Optimization (BayesOpt) is a sequential optimization strategy for finding the global maximum of black-box functions, known as objective functions \cite{Mockus1974, Mockus1977, Snoek2012}.
A benefit of using BayesOpt is that it is not necessary to know the derivative of the function that is being maximized.
BayesOpt adopts a sequential approach where all previous knowledge about the function $f(\mathbf{x})$ is used for selecting data points creating a convergence towards the global maximum.
First, a surrogate model of the function to be maximized is built, which is updated as new data points are evaluated, while also suggesting the next evaluation point.
The most popular surrogate model used in BayesOpt is the Gaussian process ($\mathcal{GP}$).
\begin{algorithm*}[ht]
    \SetAlgoLined
    Define prior bounds on function $f$ \\ 
    Observe $f$ at $n_{0}$ initial points \\ 
    \While{$n \le N$}{
        Update the posterior distribution (GP) on $f$ based on all previous data points \\
        Find the maximizer of the acquisition function, $x_{n}$, using the posterior distribution \\
        Find $f(x_{n})$ \\
        Increment $n$ \\
    }
    Return: The point with the highest evaluated $f(x)$ or the point with the largest posterior mean.
    \caption{A pseudo-code for Bayesian Optimization}
    \label{algorithm:BasicBayesOpt}
    \end{algorithm*}
    
$\mathcal{GP}$ uses the multivariate Gaussian distribution over the previous evaluated data points, described by a mean function $\mu(\mathbf{x}) = \mathbb{E}[f(\mathbf{x})]$ and a covariance function, or kernel, $k({x}, {x}') = \mathbb{E}[(f({x}) - \mu({x}))(f({x}') - \mu({x}'))]$.
The kernel used in this study was the Mat\'{e}rn kernel, whose form and details can be found in Ref. \cite{Genton2002}.
The second main component of BayesOpt is the acquisition function.
This is calculated using the $\mathcal{GP}$ and implements trade-offs between exploration of the parameter space and exploitation in BayesOpt.
The acquisition function optimizes the search for the maximum, while exploring regions where the $\mathcal{GP}$ is more uncertain.
Two common traditional types of acquisition functions are expected improvement (EI) and upper confidence-bound (UCB).
The EI acquisition function measures the expectation of improvement in the objective function based on the predicted distribution of the $\mathcal{GP}$.
When exploring the parameter in EI, points associated with high uncertainty are more likely to be chosen,  while during exploitation, parameters with high values of the mean are selected.
The next point to evaluate is chosen by $x_{n+1}=\text{max}_{x \in X}(\text{EI}(x))$ using $x^{+}$, the best parameter found so far. $\text{EI}(x)$ is defined as:
\begin{equation}
    \text{EI(x)} = f(x^{+} - \mu(x))\Phi(Z) + \sigma(x)\phi(Z)
     \label{eqn:EI}
\end{equation}
where $f(x^{+})$ has the highest value and is thereby the best observed value of the objective function, $\mu(x)$ and $\sigma(x)$ are the mean and standard deviation of the $\mathcal{GP}$ and $\phi(Z)$ and $\Phi(Z)$ are the probability and cumulative normal distributions where
\begin{equation}
    Z = \frac{f(x^{+}) - \mu(x)}{\sigma(x)}.
\end{equation}
\noindent The UCB acquisition function is defined as:
\begin{equation}
    \text{UCB}(x) = \mu(x) + \xi \sigma(x)
     \label{eqn:UCB}
\end{equation}
where $\xi \ge 0$ controls the balance between exploration of the parameter space ($\xi \sim 1$) and exploitation ($\xi \sim 0$).
This acquisition function can be described as the maximum value across all solutions of the weighted sum of the mean of the $\mathcal{GP}$ and the standard deviation of the $\mathcal{GP}$.
The next point to evaluate is chosen by the maximum of the acquisition function, $x_{n+1}=\text{max}_{x \in X}(\text{UCB}(x))$, in a similar fashion to the EI acquisition function.\\
BayesOpt has the benefit of reducing the amount of parameter space searched to find the maximum compared to other methods, such as a basic grid search, or a random search.
Since past evaluations are taken into account, BayesOpt tries to focus on the parameter space area where the maximum of the black-box function occurs while not wasting much time in those parameter space areas that are not solutions.
This greatly reduces the time required to fit complex functions, such as the GEANT4 simulations in this study.

An example that demonstrates this iterative process with BayesOpt for the function $f(x) = ( 1/(x^{2} + 1) + e^{-(x-4)^2/2})\sin(x)$ is shown in Figure \ref{fig:BayesOptExample}.
This figure starts with three random points shown as red filled markers in the top left panel.
The target function is displayed as a solid green line while the mean of the $\mathcal{GP}$ is displayed as a dashed black line.
The confidence intervals of the $\mathcal{GP}$ are shown as dark blue ($1\sigma$) and light blue ($2\sigma$) shaded regions.
An acquisition function using the upper confidence bound is shown as the solid red curve while the next point selected for evaluation is displayed as a yellow star on the acquisition function curve.
As the number of evaluations increases, the $\mathcal{GP}$ quickly improves its description of the function and hones in on the global maximum.
A basic BayesOpt pseudo-code is found in Algorithm \ref{algorithm:BasicBayesOpt}.

Bayesian Optimization has been used extensively in recent years for hyper-parameter tuning in machine learning models \cite{Snoek2012, Bergstra2011, Shahriari2016}, and also in a wide range of fields for scientific studies such as choosing experimental techniques in drug discovery \cite{Negoescu:2011}, improving quantum annealing \cite{Pelofske:2020} and tuning free-electron lasers \cite{Duris:2019}.
An example from nuclear structure theory is given in a work by Ekstr{\"o}m et al \cite{Ekstrom:2019} where it has been used to constrain the coupling constants in chiral effective field theory descriptions of the strong interaction.
In a recent work \cite{Marshall:2020}, a Bayesian Analysis was used to fit the angular distributions from a transfer reaction and extract spectroscopic factors.

In the current work, Bayesian methods incorporating GEANT4 simulations and experimental data are used to extract energies and intensities of states populated in a transfer reaction. For this work, we have chosen to use BayesOpt and the Python package \emph{BayesianOptimization} for the BayesOpt algorithm \cite{BayesOpt}.

\section{Results}
The $^{12}$C(d,p) Q-value spectrum was produced both from the experimental data and in the GEANT4 simulation, assuming the ground state and three excited states were populated. 
The GEANT4 energy loss tables were used to calculate the energy of the proton and the $^{12}$C at the center of the target for both the experimental and the simulated data.
 The experimental data were used in the Bayesian Optimization for comparison and to provide the best fit.
A recent paper \cite{Marshall:2020} shows how Bayesian optimization can be used with GEANT4 simulations to fit angular distributions from low-energy direct reactions.
We performed BayesOpt for two different acquisition functions: EI and UCB, shown in Equations \ref{eqn:EI} and \ref{eqn:UCB}, respectively. 
The inverse of $\chi^{2}$ between the experimental Q-value and that from the simulation was maximized in order to find the best (lowest) $\chi^{2}$ fit to the data.
The lowest value of $\chi^{2}$, $\chi^{2}_{min}$, was stored and used for the extraction of error bars.  
For each state, the energy was changed by a step, dE, and a Bayesian optimization was performed allowing the other states to vary until a best fit was found. The dE that produced $\chi^{2}~=~\chi^{2}_{min} + 1$ was used to define $1 \sigma$ errors.

\begin{figure}
    \centering
    \includegraphics[scale=0.3]{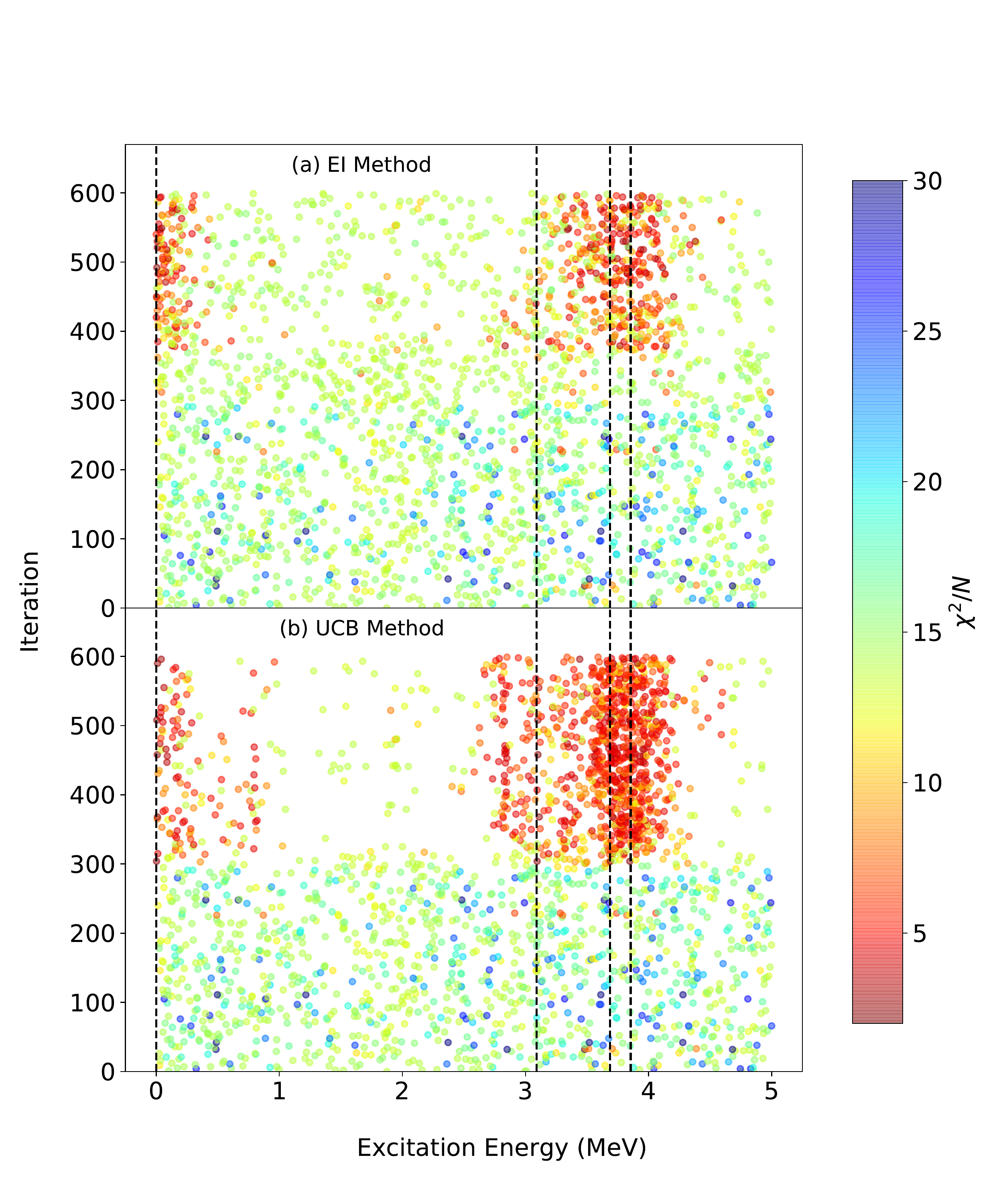}
    \caption{(Color online) Progression of the Bayesian Optimization for two acquisition functions: (a) EI and (b) UCB. 
             The BayesOpt method was used with up to four states in $^{13}$C in the region below 5 MeV in excitation energy.
             The first 300 iterations of the BayesOpt process used random iteration with an additional 300 iterations to find the optimal solution.
             For each iteration, the $\chi^{2}/N$ was calculated while the inverse was used as the objective function to be maximized.
             Plotted as vertical dashed lines are the first four states of $^{13}$C: 0.00, 3.09, 3.68 and 3.85 MeV.
             \label{fig:Model_Comparison}}
\end{figure}

To smooth out fluctuations from the simulation, 200,000 events were simulated for each iteration step of the BayesOpt process.
Since only the upstream YY1 detector was used to generate the Q-value spectrum, the simulation was set to only populate light ions uniformly in the laboratory angular range of 150$^{\circ}$-175$^{\circ}$.
This allows for a large increase in efficiency of the simulation as $\sim$35\% of the simulated events are measured in the YY1 detector.
Before the $\chi^{2}$ was calculated, the simulated data was scaled to the intensity in the experimental Q-value plot using a $\chi^{2}$ fit. 
The results using the EI acquisition function are shown in Figure \ref{fig:Model_Comparison}(a) and using the UCB acquisition function in Figure \ref{fig:Model_Comparison}(b).
For the BayesOpt process for both acquisition functions, the number of random iterations taken was set to 300.
After these random iterations, the exploitation phase of BayesOpt was initiated for a total of 300 more iterations.
The excitation energies of four states as a function of iteration number are plotted in Figure \ref{fig:Model_Comparison} along with the known values, plotted as dashed vertical lines.
The value of $\chi^{2}/N$ is denoted by color going from red (low $\chi^{2}/N$ for a good fit) to blue (high $\chi^{2}/N$ for a bad fit).
During the exploration phase for both acquisition functions, the $\chi^{2}/N$ remains mostly above 10. 
When the $\mathcal{GP}$ switches over to the exploitation function, the $\chi^{2}/N$ remains below 15 for both acquisition functions.  The line between the random exploration phase and the exploitation phase is clearly visible by an almost discrete change in color from blue to red. \\
The UCB method was more successful at converging quickly than the EI method, as illustrated by the number of low (red) $\chi^{2}/N$ points starting at 300 iterations.

\begin{figure}
    \centering
    \includegraphics[scale=0.30]{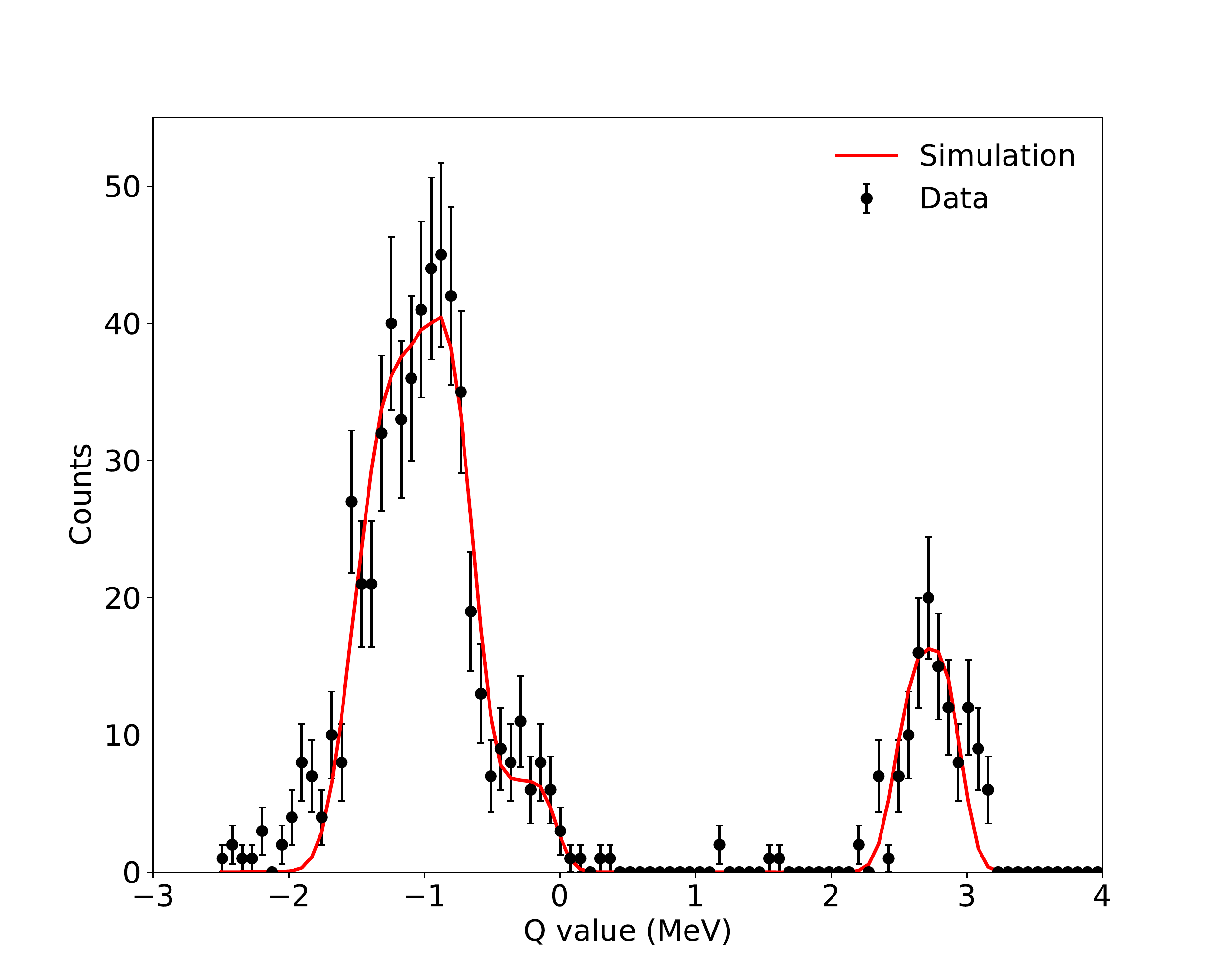}
    \caption{(Color online) Q-value spectrum comparing the best fit from 300 random exploration iterations and 300 exploitation iterations with the UCB acquisition function (red solid line) to the experimental data (black points with error bars).
             \label{fig:BayesOpt_BestFit}}
\end{figure}

The Q-value spectrum from the best fit is shown in Figure \ref{fig:BayesOpt_BestFit}.
The excitation energies from the Bayesian optimization are 3.126, 3.713, and 3.894 MeV.  
As the excited states of $^{13}$C are well known, it is possible to make meaningful comparisons, as shown in Table \ref{Table:BestFitParameters}.
There is good agreement between the fitted and known energies within error bars that are all less than 90~keV. This best fit from BayesOpt has a $\chi^{2}/N$ of 2.4.

\renewcommand{\arraystretch}{2}
\begin{table*}
    \centering
    \begin{tabular}{|c | c | c | c |}
        \hline
        BayesOpt Energy (MeV) & I/I(G.S.) & Fit (MeV) & Known Energy (MeV)\\
        \hline \hline
        $0.009^{+0.007}_{-0.009}$ & 1 & $-0.078$ & 0 \\ \hline
        $3.126^{+0.084}_{-0.060}$ & 1.115 & $2.993$ & 3.089 \\ \hline
        $3.713^{+0.057}_{-0.042}$ & 4.087 & $3.642$ & 3.685 \\ \hline
        $3.894^{+0.032}_{-0.053}$ & 2.768 & $4.112$ & 3.854 \\ \hline
    \end{tabular}
    \caption{Table of excitation energies from the BayesOpt process fitting of the $^{12}$C(d, p) reaction data compared to known values from literature.
    The energy is quoted in MeV and errors are $1\sigma$.
    I/I(G.S) is the intensity of the state relative to the intensity of the ground state.
    \label{Table:BestFitParameters}} 
\end{table*}



\section{Conclusions}
The GEANT4 toolkit with Bayesian Optimization study has been completed for the test run of the $^{12}$C(d, p) reaction.
The results obtained from BayesOpt agree with the known states in $^{13}$C, within error bars, and all within 90~keV.  
The structure of $^{13}$C is well known in the region below 5~MeV in excitation, providing a suitable benchmark for the $^{12}$Be(d,p) experiment that followed the run with $^{12}$C and populated poorly-known states in $^{13}$Be.  \\
Other techniques, such as Markov Chain Monte Carlo (MCMC), provide alternate Bayesian methods to analyze the spectrum using GEANT4.  
The MCMC method would require significant numbers of walkers and steps per walker to fit to the data, resulting in 1,000's to 10,000's of individual simulations.   In this work only 600 simulations were required.\\
BayesOpt has the benefit of reducing computation time extensively for these measurements when compared to other methods such as Markov Chain Monte Carlo and can be used to extract spectroscopic information from experimental data relevant to nuclear structure studies.
\section {Acknowledgements}
This material is based upon work was supported in part by the U.S. Department of Energy, Office of Science, Office of Nuclear Physics under Award Number DE-FG02-96ER40963 (UTK) and the U. S. National Science Foundation under Award Numbers PHY-1404218 (Rutgers) and PHY-2011890 (Notre Dame). This material is based upon work supported by the U.S. Department of Energy, Office of Science, Office of Nuclear Physics under Contract No. DE-AC05-00OR22725 (ORNL). The authors are grateful for support from NSERC, Canada Foundation for Innovation, Nova Scotia Research and Innovation Trust. The support from  RCNP, Osaka, for the target is gratefully acknowledged. TRIUMF receives funding via a contribution through the National Research Council Canada.  This work was supported by the National Research Foundation of Korea (NRF) grant funded by the Korea government (MSIT) Nos. 2020R1A2C1005981 and 2016R1A5A1013277. This work was partially supported by STFC Grant No. ST/L005743/1 (Surrey).

\bibliography{C12dp_Bayes}

\end{document}